\documentclass[twocolumn,aps,showpacs,floatfix,prc]{revtex4}

\usepackage{graphicx} 

\newcommand{\ee}{\end{equation}} 
\newcommand{\eq}{{\,=\,}} 
\newcommand{\bm}{\bf}

\begin{document} 
 
 
\title{Equation of state dependence of Mach cone like structures in Au+Au collisions}
  
\date{\today}
  
\author{Victor Roy}
\email{victor@veccal.ernet.in}
\author{A. K. Chaudhuri} 
\email{akc@veccal.ernet.in} 
\affiliation{Variable Energy Cyclotron Centre, 1-AF, Bidhan Nagar,
Kolkata - 700 064, India} 
  
\begin{abstract} 
In jet quenching, a hard QCD parton, before fragmenting into a jet of hadrons, deposits a fraction of its energy in the medium. As the parton moves nearly with speed of light, much greater that the speed of sound of the medium, quenching jet can generate Mach shock wave. We have examined the possibility of Mach shock wave formation due to jet quenching.
Assuming that the   deposited  energy quickly thermalize, we simulate the  hydrodynamic 
evolution of the QGP fluid with a quenching jet and subsequent particle production. Angular distribution of pions, averaged over all the jet trajectories, resembles 'conical flow' due to Mach shock wave formation. 
 However, speed of sound dependence  of the simulated Mach angles are at variance with that due to shock wave formation in a static medium or in a medium with finite velocity.
   
 \end{abstract} 
 
 
\maketitle 

\section{Introduction} \label{sec1} 
   From the general theoretical grounds, it was
predicted \cite{QGP3jetqu} that in a dense, deconfined medium, high-speed partons will suffer energy loss, significantly modifying the fragmentation function, which in turn will lead to suppressed production of hadrons. The phenomena called jet quenching, was later verified at Relativistic Heavy Ion Collider (RHIC), in Au+Au collisions at $\sqrt{s}_{NN}$=200 GeV \cite{Adcox:2001jp,Adler:2002xw,Adams:2003kv}. While jet quenching is verified,   how the lost energy is transported in the dense medium is uncertain.  
It has been suggested that a fraction of lost energy will go to collective excitation, call the
"conical flow" \cite{Stoecker:2004qu,shuryak,Casalderrey-Solana:2005rf}. The parton moves with speed of light ($c_{jet}\approx 1$), greater
than the speed of sound of the medium ($c_s$), and the quenching jet can produce a shock wave with Mach cone angle, $\theta_M=cos^{-1}\frac{c_s}{c_{jet}}$. Resulting conical flow will have characteristic peaks at $\phi=\pi-\theta_M$ and $\phi=\pi + \theta_M$. Both in STAR \cite{Wang:2004kf} and PHENIX \cite{Jacak:2005af} experiments,  indication of such peaks are seen in azimuthal distribution of secondaries associated with high $p_T$ trigger in central Au+Au collisions. As Mach cone is sensitive to the speed
of sound of the medium, it raises the possibility of measuring
the speed of sound of the deconfined matter of Quark-Gluon-Plasma.
Mach like structure (splitting of away side peak) can also
be obtained in various other models, e.g. gluon Cerenkov like radiation models \cite{Dremin:2005an,Koch:2005sx}, the parton cascade model \cite{Ma:2006mz}. Recently in  \cite{Chesler:2007an} energy density wake produced by a heavy quark moving through 
a strongly coupled N=4 supersymmetric, Yang-Mills plasma is computed using ADS/CFT correspondence. Mach  like structures is also observed for quark velocity greater than the speed of sound of the medium.

Mach like structures are best explored in hydrodynamics models. Several authors \cite{Renk:2005si,Ruppert:2007mm,Torrieri:2009mv,Betz:2008js,Noronha:2008un,Betz:2008wy,Neufeld:2008fi,Chaudhuri:2005vc,Chaudhuri:2006qk,Chaudhuri:2008zz,Chaudhuri:2007vc,Chaudhuri:2007gq} have explored the possibility of Mach shock wave formation in a  jet event. 
 
In order to understand the dynamics of energy transport by the QGP fluid, in \cite{Chaudhuri:2005vc,Chaudhuri:2006qk,Chaudhuri:2008zz,Chaudhuri:2007vc,Chaudhuri:2007gq},
we have modeled the quenching jet as a (time dependent) source and
solved  hydrodynamical energy-momentum conservation equations. It is implicitly assumed that the energy deposited locally along the trajectory of the jet is quickly thermalized and evolve hydrodynamically. 
Explicit simulations of the "`Hydro+Jet"' model \cite{Chaudhuri:2006qk}, indicate that in a single jet event, azimuthal distribution of pions do not resemble the structure characteristic of 'conical flow'. Instead of two peak distribution as expected in a conical flow,    the   distribution exhibits a single peak structure. The peak position
depend  on the jet trajectory. However, when averaged over all the jet trajectories
(di-jet can be produced anywhere on the surface of the fireball and since
all positions are equally likely, one needs to average over all possible di-jet production points), hydrodynamical evolution with a quenching jet produces a 
angular distribution characteristic of 'conical flow, i.e. two peaks separated by a dip at $\phi=\pi$  \cite{Chaudhuri:2006qk}.  When simulated with realistic initial condition, equation of state etc, azimuthal distribution of excess pions,  averaged over all possible jet trajectories, reasonably well reproduce the STAR and PHENIX data \cite{Wang:2004kf} on the dijet-hadron correlation in 0-5\% centrality Au+Au collisions \cite{Chaudhuri:2006qk,Chaudhuri:2008zz,Chaudhuri:2007vc,Chaudhuri:2007gq}.
However, 
it is yet uncertain, whether or not the observed dijet-hadron correlation  is due to Mach shock wave formation. In an evolving medium, due to finite fluid velocity, Mach shock fronts are distorted \cite{Satarov:2005mv}.  
Model calculations \cite{Satarov:2005mv} indicate that   the inside shock front is pushed in and the outside shock front is pushed out. If $\theta_M^+$ and $\theta_M^-$ denote the outside and inside shock cone angles, then for fluid velocity $u$, the Mach angles are changed as \cite{Satarov:2005mv},

\begin{equation} \label{eq9}
\theta_M^\pm \approx \frac{\pi}{2}-tan^{-1}\left[\gamma_u \frac{\gamma_s c_s \pm \gamma_u u}{1 \mp \gamma_s c_s \gamma_u u}\right]
\end{equation}

\noindent 
where $\gamma_u=(1-u^2)^{-1/2}$, $\gamma_s=(1-c_s^2)^{-1/2}$.  
For supersonic flow i.e. at $u > c_s$, the inside cone angle $\theta_M^- > \frac{\pi}{2}$ and jet
trajectory lies outside the Mach region. Then depending upon the fluid velocity,
inside shock front may be obliterated and only the outside shock front survives.
Angular distribution due to distorted shock fronts may  lead to single peaked angular distribution,   as obtained  in simulations in the 'Hydro+Jet' model \cite{Chaudhuri:2006qk}. On averaging over different jet trajectories, two peak structures can still be obtained. However, in that case, peak positions will corresponds to $\phi=\pi/2 \pm \theta_M^+$ and simulated Mach angles will show definite dependence on speed of sound of the medium, decreasing with the speed of sound. 

 In the present paper, in the "`Hydro+Jet"' model \cite{Chaudhuri:2005vc}, we have investigated the speed of sound dependence of the
 azimuthal distribution of secondaries.    For a simple equation of state, $p=c_s^2 \varepsilon$,  azimuthal distribution of $\pi^-$ is studied as a function of speed of sound, $c_s^2$=0.1-0.4.  It is shown that the   equation of state dependence of the azimuthal distribution of $\pi^-$  is unlike that due to a Mach shock wave in a static or moving medium.  
 
The paper is organized as follows: in section \ref{sec2}, we describe the model used to study jet quenching in an expanding medium. In section \ref{sec3}, simulation results are discussed. Finally, summary and conclusions are given in  section \ref{sec4}.

 \begin{figure}[t]
 \center
 \resizebox{0.35\textwidth}{!}
 {\includegraphics{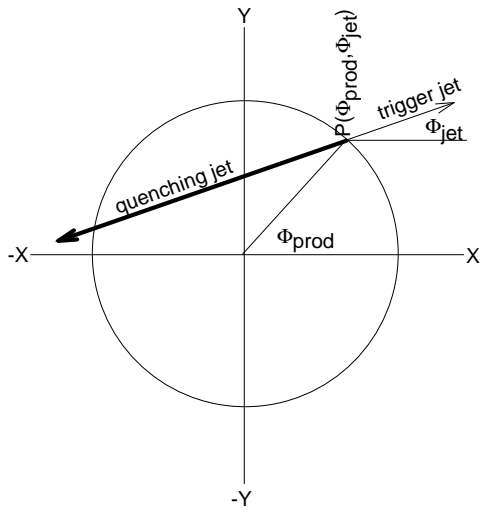}}
\caption{ schematic representation of a jet moving through the
medium. The high $p_T$ pair is assumed to produce at $P$ on the surface of the fireball characterized by the angle
$\phi_{prod}$. One of the jet escapes forming the trigger jet, the
other move in the fireball at an angle $\phi_{jet}$.}\label{F1}
\end{figure}

\section{Hydro+Jet model} \label{sec2}

A schematic representation
of the jet moving through the medium is shown in Fig.\ref{F1}.
We assume that just before hydrodynamics become applicable, a pair 
of high-$p_T$ partons is produced.  
Strong jet quenching and survival of the trigger jet, forbids production
in the interior of the fireball. Jet pairs can be produced only on a thin shell on the surface of the fireball. Without any loss of
generality, we assume that the pair is produced on the boundary of the fireball. 
For Au+Au collisions at impact parameter $b$,
the fireball boundary is an ellipsoid with minor and major axis,
$A=R-b/2$ and $B=R \sqrt{1-b^2/4R^2}$, 
with R=6.4 fm.
As shown in Fig.\ref{F1}, the jet production point can be characterized by the angle $\phi_{prod}$ 
($0 \leq \phi_{prod} \leq 2\pi$) only .
One of jet moves outward  and escapes, forming the trigger jet. The other enters into the fireball. The trajectory of the jet can be designated by the angle $\phi_{jet}$,
($-\frac{\pi}{2} \leq \phi_{jet} \leq +\frac{\pi}{2}$).
The fireball is expanding and cooling. The ingoing parton travels 
at the speed of light and loses energy in the fireball.
We note that just at the
creation point $P$, the jet is a hard "undressed" parton. The "undressed" parton loses energy by constantly emitting radiation (gluons) which in turn emit new radiation, developing a  shower. The shower has a complicated structure, but multiplicity in the shower grows non-linearly and very soon the region near the head of the jet
can become a "macroscopic" body providing large perturbation
to the medium. We treat this perturbation with a simple assumption. We assume that the perturbation is thermalized
and acts as a source of energy and momentum to the fireball.
We solve the energy-momentum conservation equation,

\begin{equation} \label{1}
\partial_\mu T^{\mu\nu}=J^\nu,
\end{equation}

\noindent where the source is modeled as,

\begin{eqnarray} 
\label{2} 
 &&J^\nu(x)=J(x)\,\bigl(1,-cos(\phi_{jet}),-sin(\phi_{jet}),0\bigr),\\
\label{3}
 &&J(x) = \frac{dE}{dx}(x)\, \left|\frac{dx_{\rm jet}}{dt}\right| 
          \delta^3(\bm{r}-\bm{r}_{\rm jet}(t)).
\end{eqnarray} 
Massless partons have light-like 4-momentum, so the current $J^\nu$
describing the 4-momentum lost and deposited in the medium by the 
fast parton is taken to be light-like, too. $\bm{r}_{\rm jet}(t)$ is 
the trajectory of the jet moving with speed $|dx_{\rm jet}/dt|\eq{c}$.
$\frac{dE}{dx}(x)$ is the energy loss rate of the parton as it moves 
through the liquid. It depends on the fluid's local rest 
frame particle density and we parameterise it as, 
\begin{equation}
\label{4}
  \frac{dE}{dx} = \frac{s(x)}{s_0} \left.\frac{dE}{dx}\right|_0
\end{equation}
where $s(x)$ is the local entropy density without the jet.    
The measured suppression of high-$p_T$ particle production in Au+Au 
collisions at RHIC was shown to be consistent with a parton energy 
loss of $\left.\frac{dE}{dx}\right|_0\eq14$\,GeV/fm at a reference 
entropy density of $s_0\eq140$\,fm$^{-3}$ \cite{Eloss}. In \cite{Chaudhuri:2007vc}, it was shown that with realistic initial conditions and equation of state, hydrodynamic evolution with parton energy 
loss of $\left.\frac{dE}{dx}\right|_0\eq14$\,GeV/fm at a reference 
entropy density of $s_0\eq140$\,fm$^{-3}$ reasonably explain the dijet-hadron correlation. Presently, we are interested in the equation of state dependence on Mach cone like surfaces. We have considered simple equation of state $p=c_s^2 \varepsilon$, with $c^2_s$=0.1,0.2,0.3 and 0.4. For the present study, with a view to enhance the effect of jet quenching, we use, $\left.\frac{dE}{dx}\right|_0\eq 28$\,GeV/fm at a reference 
entropy density of $s_0\eq140$\,fm$^{-3}$.
Reference energy loss $\left.\frac{dE}{dx}\right|_0\eq 28$\,GeV/fm   may seem to high, but energy deposited to the medium is never too high due to the factor $s(x)/(s_0=140)$ in Eq.\ref{4}. 
The di-jet is produced at the peripheri where local entropy 
density $s(x) << s_0$ and at early time the energy loss is not large. By the time, jet has reached the centre, local entropy density is degraded very much. For example, in boost-invariant longitudinal motion (
in Bjorken dynamics), for a jet moving along the diameter, by the time jet reaches the centre, local entropy density
is degraded by a factor of ~10 and again the jet energy loss is not large. 
Indeed, we have checked that average energy loss is $\sim$ 10-15 GeV, depending on the jet trajectory. 
It may
be mentioned that even though in eq.\ref{4}, partonic energy loss is assumed to depend only on the fluid rest frame density, it could as well depend on the  parton energy \cite{Baier:1996kr,Gyulassy:2003mc,Accardi:2004gp}.

For the hydrodynamic evolution we use a modified version of the publicly available hydrodynamic code AZHYDRO \cite{QGP3v2,AZHYDRO}. The code
  is formulated in 
$(\tau,x,y,\eta)$ coordinates, where $\tau{=}\sqrt{t^2{-}z^2}$ is the
longitudinal proper time, 
$\eta{=}\frac{1}{2}\ln\left[\frac{t{+}z}{t{-}z}\right]$ 
is space-time
rapidity, and $\bm{r}_\perp{\,=\,}(x,y)$ defines the plane transverse to the 
beam direction $z$. AZHYDRO employs longitudinal boost invariance 
along $z$ but this is violated by the source term (\ref{3}). We 
therefore modify the latter by replacing the $\delta$-function 
in (\ref{3}) by

\begin{eqnarray}
\label{5}
  \delta^3(\bm{r}-\bm{r}_{\rm jet}(t)) &\longrightarrow&
  \frac{1}{\tau}\,\delta(x-x_{\rm jet}(\tau))\,\delta(y-y_{\rm jet}(\tau))
\nonumber\\
 &\longrightarrow&\frac{1}{\tau} \, 
\frac{e^{-(\bm{r}_\perp-\bm{r}_{\perp,{\rm jet}}(\tau))^2/(2\sigma^2)}}
      {2\pi\sigma^2}
\end{eqnarray}
with $\sigma{\,=\,}0.35$\,fm. 

Intuitively, this replaces the 'needle'
(jet) pushing through the medium at one point by a 'knife' cutting the
medium along its entire length along the beam direction. If we disregard the radial motion, the motion of a pointlike source produces a 'conical' wave but motion of a 'linelike' source will produce a 'wedge like' flow. In a 'wedge like' flow, effect of jet quenching will be enhanced, the two peak angular structure will be more pronounced,
than in a 'conical' flow. However, the peak positions themselves (or the cone angle) will remain unchanged. 
In the present study, we are interested in the equation of state dependence of Mach cone like surfaces produced by a quenching jet. 
While a complete study of jet quenching do require a full 
(3+1)-dimensional hydrodynamic calculation, qualitative change in Mach cone like surfaces due to different equation of states can be adequately addressed in a
boost-invariant simulation. We have tried to estimate the enahancement expected due to the assumption of boost invariance.
 PHOBOS collaboration has measured the pseudo-rapidity density for the charged particles  \cite{Back:2002wb}. Over a pseudo-rapidity range $\sim$ (-6  to 6) boost-invarinace is approximately valid in the range $\sim$ (-2.5 to + 2.5). The assumption boost-invariance then overestimate the particle yield by a factor $\sim$1.5. Effect of jet quenching will be overestimated by a similar factor. In less central collisions, the effect will be still less overestiamted. 
It may also be mentioned here that the Gaussian like energy loss distribution (see Eq.\ref{5}) is  at variance with realistic calculations.  Microscopic studies \cite{Yarom:2007ni,Chesler:2007sv,Neufeld:2008hs} do indicate that the parton  energy loss is more like a 'spike' at short distances and an exponential fall off at large distances.   
 
The modified hydrodynamic equations in $(\tau,x,y,\eta)$ coordinates
read \cite{Chaudhuri:2005vc}

   \begin{figure}[t]
 \center
 \resizebox{0.5\textwidth}{!}
 {\includegraphics{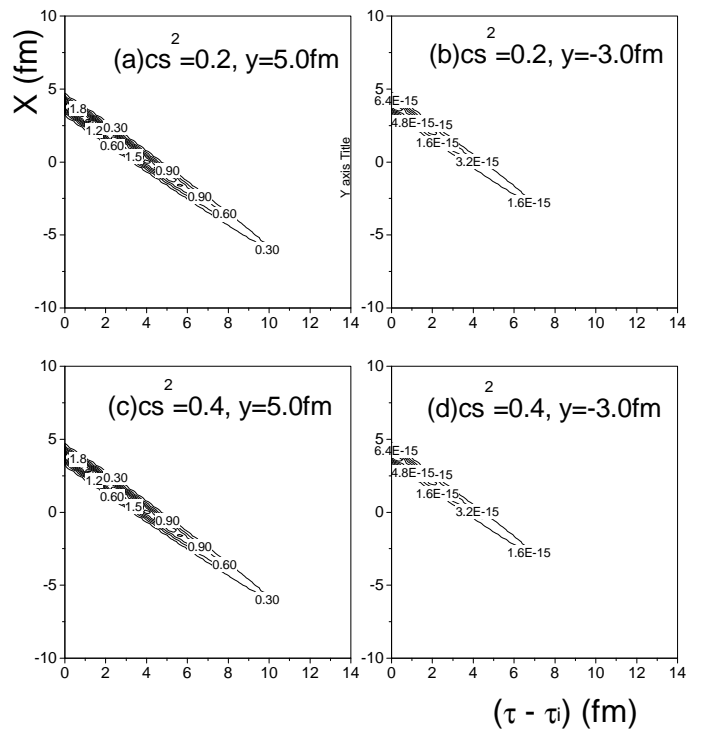}}
\caption{Contour plot of energy deposition in $x-\tau$ plane, by a quenching jet in medium with square speed of sound $c_s^2$=0.2 (panels (a) and (b)) and $c_s^2$=0.4 (panels (c) and (d)). The quenching jet is produced at (x=4.5 fm, y=5 fm) and moving parallel to the X-axis. Energy deposition at fixed y=4.5 fm and y=-3 fm are shown.   Note that away from the trajectory, the quenching jet deposit negligible energy in to the medium.} \label{F2}
\end{figure}

%
\begin{eqnarray} 
\label{6} 
  \partial_\tau \tilde{T}^{\tau \tau} + 
  \partial_x(\tilde{v}_x \tilde{T}^{\tau \tau}) +
  \partial_y(\tilde{v}_y \tilde{T}^{\tau \tau}) 
  &=& - p + \tilde{J},
\\ 
\label{7} 
  \partial_\tau \tilde{T}^{\tau x} +
  \partial_x(v_x \tilde{T}^{\tau x}) +
  \partial_y(v_y \tilde{T}^{\tau x}) 
  &=& - \partial_x \tilde{p} - \tilde{J}, \quad
\\ 
\label{8} 
  \partial_\tau \tilde{T}^{\tau y} +
  \partial_x(v_x \tilde{T}^{\tau y}) +
  \partial_y(v_y \tilde{T}^{\tau y}) 
  &=& -\partial_y \tilde{p},  \quad
\end{eqnarray}  
%
where $\tilde{T}^{\mu\nu}\eq\tau T^{\mu\nu}$, 
$\tilde{v}_i{\eq}T^{\tau i}/T^{\tau\tau}$,
$\tilde p\eq\tau p$, and $\tilde{J}\eq\tau J$.

  To simulate central Au+Au collisions at RHIC, we use the standard 
initialization described in \cite{QGP3v2} and provided in the 
downloaded AZHYDRO input file \cite{AZHYDRO}, corresponding to a 
peak initial energy density of $\varepsilon_i \eq 30$ $GeV/fm^3$ at
$\tau_i \eq 0.6$ $fm/c$. 
Since we are exploring the equation of state dependence on the splitting of the away side jet, we assume a simple equation of state, $p(\varepsilon)=c^2_s \varepsilon$, with $\varepsilon(T)=\frac{\pi^2}{30} g_q T^4$, $g_q=47.5$. We also neglect the possibility of phase transition. Till the freeze-out the medium evolve as QGP. We assume instantaneous hadronisation at the freeze-out temperature   $T_F$=100 MeV and using Cooper-Frye prescription compute invariant distribution of $\pi^-$. It must be mentioned that   
in reality, QGP fluid will cool with time and at a critical temperature ($T_c$) will undergo a phase transition to hadronic matter. However, speed of sound dependence of Mach cone will then be more
complicated. 
If $\tau_i$ and $\tau_F$ are the initial and freeze-out time then in an evolving medium the Mach cone angle will be defined as 

\begin{equation}
cos \theta_M=\frac{1}{\tau_F-\tau_i}\int^{\tau_F}_{\tau_i} c_s(\tau) d\tau
\end{equation} 

If $c_s$ change between $\tau_i$ and $\tau_F$, unraveling speed of sound dependence of the  Mach like structures will be more complicated.

\section{Results} \label{sec3}

In the following, we will simulate Au+Au collisions at zero impact parameter.
In b=0 Au+Au collision, the reaction zone is spherically symmetric. For a spherically symmetric fireball, the fireball can be rotated by the angle $\phi_{jet}$
to make  $\phi_{jet}$=0 and the jet trajectory can be characterized by the angle $\phi_{prod}$ only, $\phi_{prod}$ varying between $[-\pi/2,\pi/2]$.   
 
 \begin{figure}[t]
 \center
 \resizebox{0.35\textwidth}{!}
 {\includegraphics{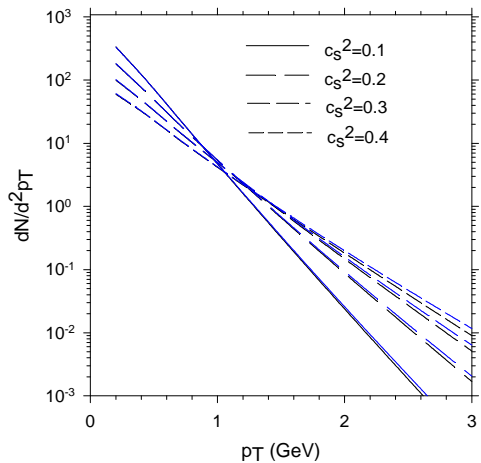}}
\caption{(color online) The black lines are transverse momentum spectrum of $\pi^-$
from evolution with out a quenching jet. The blue lines are from evolution with a quenching jet.}
 \label{F3}
\end{figure}  

Let us discuss speed of sound dependence of energy deposition by a quenching jet.
In Fig.\ref{F2}, we have shown the contour plot of energy deposition by a quenching jet in   medium
with square speed of sound $c_s^2$=0.2 and 0.4.  The quenching jet is produced at (x=4.5 fm, y=5 fm) and is moving inward parallel to  the x-axis, and the contours are drawn in the $x-\tau$ plane
 at a fixed y=5 fm (panels (a) and (c)) and y=-3 fm (panels (b) and (d)).
 With time,  the quenching jet deposits less and less energy. It is understood. Jet energy loss is weighted by the entropy density. As the fluid evolves, entropy density drops and the jet deposit less and less energy to the medium.  One also notice that away from the jet trajectory, the jet deposit hardly any energy in the medium. For example, energy deposition falls by a factor of $\sim 10^{-15}$ from y=5 fm to y=-3 fm. It is interesting to note that the perturbation to the medium with $c_s^2$=0.2 or 0.4, by the quenching jet, is approximately same. The quenching jet deposit approximately same energy  to the medium.  Other conditions remaining the same, low $c_s$ medium  evolve for longer duration than in  high $c_s$ medium. However,   the energy transfer from jet to medium is inefficient at late time and approximately same energy is deposited in the medium. 

It is instructive to compare transverse momentum distribution of secondaries from evolution of fluid with and with out a quenching jet.
The black solid, long dashed, medium dashed and short dashed lines in Fig.\ref{F3}, are the transverse momentum distribution of $\pi^-$, from   fluid with square speed of sound, $c_s^2$=0.1, 0.2, 0.3 and 0.4 respectively. Other conditions remaining the same, particle production depend considerably on the speed of sound. With increasing speed of sound, 
$p_T$ spectra flattens, but yield at low $p_T$ decreases. Freeze-out occur early in high $c_s$ medium than in a low $c_s$ medium. Flattening of the $p_T$ spectra is indicative of early freeze-out. If $p_T$ spectra  is approximated as $dN/d^2p_T \propto exp(-M_T/T)$, effective source temperature increases 'linearly' with speed of sound. 
In Fig.\ref{F3}, the blue lines are transverse momentum distribution of $\pi^-$
with a quenching jet. The jet production angle is  $\phi_{prod}$=1.05 rad. Effect of jet quenching is not large on the transverse momentum distribution. Yield at high $p_T$  is increased by 10-20\%. At low $p_T$, effect of the quenching jet is even less.  

   \begin{figure}[t]
 \center
 \resizebox{0.35\textwidth}{!}
 {\includegraphics{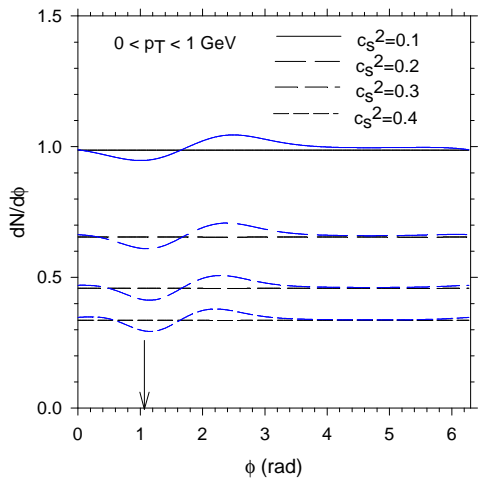}}
\caption{(color online) The black lines are azimuthal distribution pions from evolution of QGP fluid with square speed of sound $c_s^2$=0.1,0.2,0.3 and 0.4. The blue lines are same with a quenching jet. The jet is produced at $\phi_{prod}$=1.05 rad, indicated by the arrow.}
 \label{F4}
\end{figure}

\begin{figure}[t]
 \center
 \resizebox{0.35\textwidth}{!}
 {\includegraphics{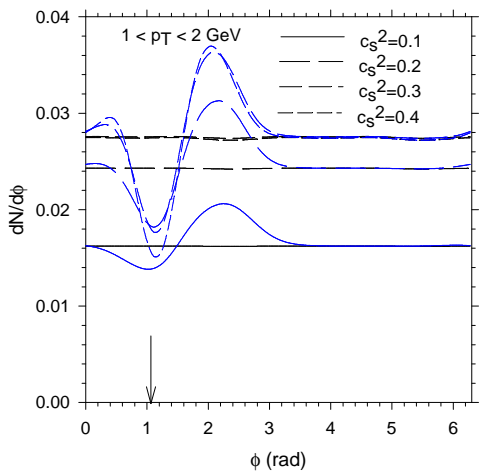}}
\caption{same as in Fig.\ref{F4} but in the $p_T$ range $1 GeV \leq p_T \leq 2 GeV$.}
 \label{F5}
\end{figure}

\begin{figure}[t]
 \vspace{0.5cm}
 \center
 \resizebox{0.35\textwidth}{!}
 {\includegraphics{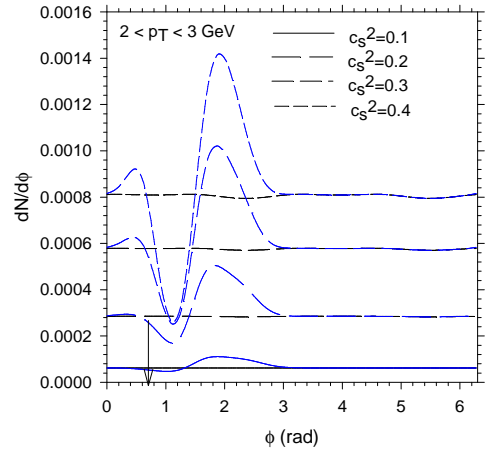}}
\caption{same as in Fig.\ref{F4} but in the $p_T$ range $2 GeV \leq p_T \leq 3 GeV$. .}
 \label{F6}
\end{figure}

Effect of jet quenching is better observed in azimuthal distribution of $\pi^-$. 
In Fig.\ref{F4}, azimuthal distribution of $\pi^-$ ($0 \leq p_T \leq 1 GeV$)  from evolution of fluid with and with out any quenching jet are shown. 
The  black solid, long dashed, medium dashed and short dashed lines are azimuthal distribution of $\pi^-$ form evolution of fluid with square speed of sound  $c_s^2$=0.1, 0.2, 0.3 and 0.4 respectively. The blue lines are same from fluid evolution with a quenching jet. Without any quenching jet, the pion distribution is flat. In b=0 Au+Au collisions, reaction zone is azimuthally symmetric and so is the azimuthal distribution of  $\pi^-$.  With a quenching jet,   $\pi^-$ distribution is modified.  The quenching jet defines a direction and azimuthal symmetry is lost. Pion distribution is no longer flat but shows a single peak structure.   
 The peak positions appear to depend marginally on the speed of sound.   We also note that in fluid evolution with a quenching jet,  $\pi^-$ production is
depleted around $\phi\approx$ 1 rad. In Fig.\ref{F4}, the arrow marked the jet production angle $\phi_{prod}$. $\pi^-$ production is depleted  approximately at the entry point of the jet. The azimuthal distribution of $\pi^-$ with a quenching jet is unlike that expected in conical flow due to shock wave. In conical flow,   azimuthal distribution is expected to be double peaked with a dip at $\phi=\pi$. However, as seen here, distribution is hardly modified at $\phi >\pi$. Also, variation of peak angle with $c_s$ is much less and opposite to that expected in a Mach shock wave.  
In Fig.\ref{F5} and \ref{F6}, azimuthal distribution of $\pi^-$ in $p_T$ range, $1 GeV \leq p_T \leq 2 GeV$ and
$2GeV\leq p_T \leq 3GeV$ are shown. Qualitatively, the azimuthal distribution of high $p_T$ pions are similar to that of low $p_T$ pions.
At  high $p_T$ also, azimuthal distribution is single peaked, associated with depleted production around
the jet production angle. However, the peaks are better defined at high $p_T$.  
 
\begin{figure}[h]
 \vspace{0.5cm}
 \center
 \resizebox{0.35\textwidth}{!}
 {\includegraphics{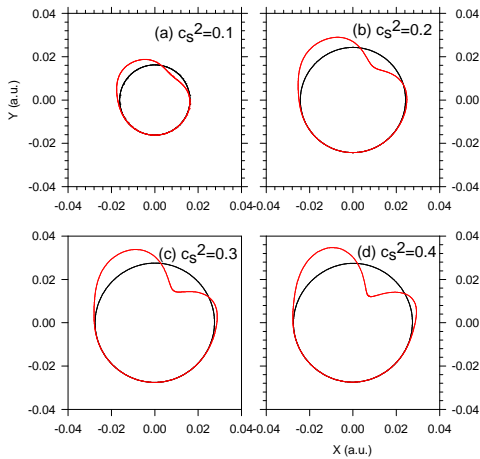}}
\caption{(color online) Azimuthal distribution of $\pi^-$ is mapped into x-y plane. The black and red lines in panel (a), (b), (c) and (d) corresponds to fluid evolution without any quenching jet and with a quenching jet produced at $\phi$=0.78 rad.}
 \label{F7}
\end{figure}

In Fig,\ref{F7}, we have mapped the azimuthal distribution of $\pi^-$ ($1 \leq p_T \leq 2$ GeV) in x-y plane. Remember that in a dynamic model like hydrodynamics, the particles are continuously emitted from a dynamic freeze-out surface. The mapping $x=\frac{dN}{d\phi}cos(\phi)$, $y=\frac{dN}{d\phi}sin(\phi)$ give 
the shape of the effective freeze-out surface. 
Black lines in Fig.\ref{F7} correspond to fluid evolution without any quenching jet. The freeze-out surface is a circle then. Effective freeze-out surface is modified with a quenching jet. The modification is   qualitatively similar for all values of $c_s$. The surface, in the lower half plane is largely unaltered. Surface on the upper half plane is modified; around the jet entry point, the surface is pushed in. As a consequence, the surface is pushed out in another direction. Apparently the matter behaves as an incompressible fluid. Modification of surface is also great when speed of sound of the medium is large. However, the modification is unlike that expected in a Mach  shock wave formation. The simmulation studies do indicate that in a single jet event, quenching jet do not lead to a double peak structure in azimuthal correlation. The result is in agreement wit  recent studies of azimuthal correlation induced by energy loss by a heavy quark \cite{Betz:2008wy}. It was  shown that a pQCD source term for heavy quark energy loss do not lead to double peak structure in azimuthal correlation.
 
Even though, in a single jet trajectory,   azimuthal distribution of $\pi^-$ do not show a double peaked structure as in a conical flow, when averaged over the jet trajectories, a double peaked structure, mimicking the conical flow, can emerge \cite{Chaudhuri:2006qk,Chaudhuri:2007gq}. A di-jet can be produced anywhere on the surface of the fireball and one need to average over all the jet trajectories. As demonstrated in \cite{Chaudhuri:2006qk}, a quenching jet in the upper half plane
 produces excess pions in the angular range $0 \leq \phi \leq \pi$, while a jet in the lower half plane produces excess pions in the angular range  $\pi \leq \phi \leq 2\pi$ and on averaging produces a double peaked azimuthal distribution with a dip at $\phi=\pi$. 

  \begin{figure}[t]
 \center
 \resizebox{0.35\textwidth}{!}
 {\includegraphics{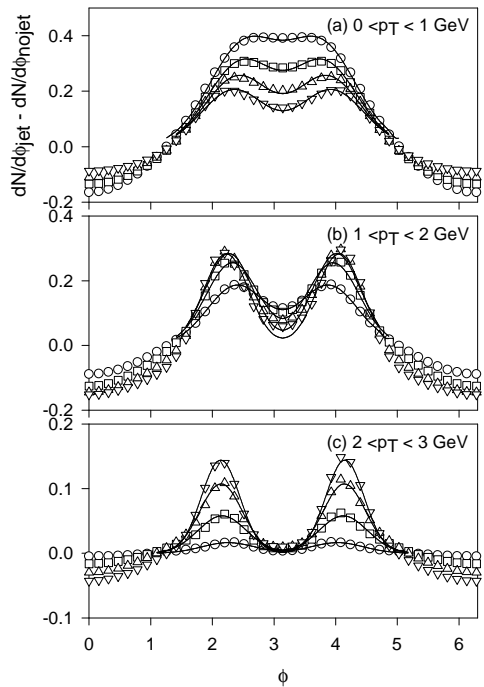}}
\caption{Jet trajectory averaged azimuthal distribution of excess pions $\left[\frac{dN}{d\phi}|_{jet}-\frac{dN}{d\phi}|_{nojet}\right]$, in the $p_T$ range, (a) $0 \leq p_T \leq 1$ GeV, (b) $1 GeV \leq p_T \leq 2 GeV$ and (c) $2 GeV \leq p_T \leq 3 GeV$. The blank circles, squares, up triangles and down triangles corresponds to square speed of sound $c_s^2$=0.1, 0.2, 0.3 and 0.4 respectively.  The lines are fit to the simulated azimuthal distribution by two Gaussians (Eq.\ref{eq10}).}
  \label{F8}
\end{figure}

  In Fig.\ref{F8},   we have shown the azimuthal distribution of 'jet trajectory averaged' excess pions [$\left<\frac{dN}{d\phi}_{jet}\right>_{av} - \frac{dN}{d\phi}_{nojet}$]. Trajectory averaged $\pi^-$ distribution   is obtained as,

\begin{equation}
\left<\frac{dN}{d\phi}_{jet}\right>_{av}=\frac{1}{\pi} \int_{-\pi/2}^{\pi/2} \frac{dN(\phi_{prod})}{d\phi} d\phi_{prod}
 \end{equation}

Excess pion distribution  
  in the $p_T$ range,  (a) $0  \leq p_T \leq 1 GeV$,
 (b) $1 GeV \leq p_T \leq 2 GeV$ and (c) $2GeV \leq p_T \leq 3 GeV$ are shown in three panels of Fig.\ref{F8}. 
Jet trajectory averaged azimuthal distribution clearly show a two peak structure   with a dip at $\phi=\pi$. At high $p_T$ and also in medium with high speed of sound, the peak structures get better defined and the dip at $\phi$=$\pi$ deepens.     We also notice that in the simulation, due to quenching jet, particle production is depleted around $\phi$=0, i.e. around the trigger jet. Depletion     around trigger jet is not observed experimentally.   We have neglected the trigger jet. The trigger jet fragment  in the vacuum and populate phase space around $\phi=0$.  Particles from evolution of the fireball and particles 
from trigger jet fragmentation can not be distinguished and depletion of particle production around $\phi=0$ is not observed.

Azimuthal distribution of trajectory averaged excess pion distribution resemble a conical flow, i.e. two peak distribution with a dip at $\phi=\pi$.  As argued earlier, if the peaks are due to distorted shock waves
in a moving medium, such that inside shock front is obliterated and only the outside shock front survives, then the azimuthal distribution of excess pions with have peaks at $\phi_{peak}=\pi \pm \theta_M^+$. Peak position or the outside shock cone angle $\theta_M^+$ will show definite variation with speed of sound (see Eq.\ref{eq9}). 
To test this,
we have fitted the azimuthal distribution of excess pions by a double Gaussian,

  \begin{figure}[t]
 \vspace{0.5cm}
 \center
 \resizebox{0.35\textwidth}{!}
 {\includegraphics{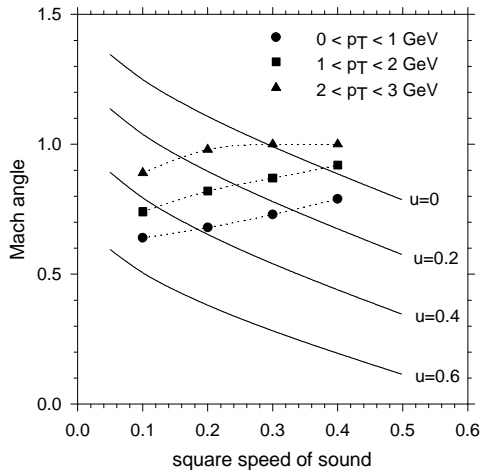}}
\caption{The solid lines (from top to bottom) are theoretical estimate of Mach angle $\theta_M^+$ [see Eq.\ref{eq9}], as a function of the square speed of sound, for fluid velocity $u$=0, 0.2, 0.4 and 0.6 respectively. Simulated Mach angles (${\theta^+_M}^{sim}$) from jet trajectory averaged $\pi^-$ distribution, in the $p_T$ range,  $0 \leq p_T \leq 1$ GeV, $1 \leq p_T \leq 2$ GeV and $2 \leq p_T \leq 3$ GeV are shown as filled circles, squares and triangles.}
  \label{F9}  
  \end{figure}
  
\begin{equation}\label{eq10}
\frac{dN}{d\phi} \propto e^{-\frac{(x- \pi- {\theta^+_M}^{sim})^2}{2\sigma^2}}  
+ e^{-\frac{(x- \pi + {\theta^+_M}^{sim})^2}{2\sigma^2} }
\end{equation}

 Fits obtained to the simulated $\pi^-$ distribution is shown in Fig.\ref{F8}.
Double Gaussian fits the azimuthal distribution of the trajectory averaged excess pions due to jet quenching. 
In Fig.\ref{F9}, we have compared the simulated Mach angles (${\theta^+_M}^{sim}$) with   Mach angles  (Eq.\ref{eq9}) in a moving medium.  The solid lines
in Fig.\ref{F9} are Mach angles in a moving medium, as a function of square speed of sound. We have shown the angles for fluid velocity, $u$=0, 0.2, 0.4 and 0.6. 
 Mach angles in a moving medium decreases with speed of sound and also with increasing fluid velocity.
The filled circles, squares and triangles in  Fig.\ref{F9} are simulated Mach angles ${\theta^+_M}^{sim}$ from the excess  pion distributions in the $p_T$ range $0 \leq p_T \leq 1GeV$,
$1 \leq p_T \leq 2GeV$ and $2 \leq p_T \leq 3GeV$ respectively. Unlike in theory,
simulated Mach angles depend on $p_T$. Speed of sound dependence of ${\theta^+_M}^{sim}$  is also inconsistent with the speed of sound dependence of   Mach angles in a moving medium.   
Simulated Mach angles will be consistent with Mach angles in a moving medium,  if, on the average, fluid velocity is less in medium with high speed of sound. 
For example,  if fluid velocity in medium with $c_s^2$=0.1, 0.2, 0.3 and 0.4 are $\sim$ 0.5, 0.38, 0.25 and 0.1 respectively, simulated Mach angles in the $p_T$ range $0 \leq p_T \leq 1 GeV$, will be consistent with the  Mach angles in a moving medium.  However, in
hydrodynamic evolution, fluid velocity grow faster in  a high $c_s$ medium than in a low $c_s$ medium. Consistency condition i.e. fluid velocity is less in high $c_s$ medium than in low $c_s$ medium can not be satisfied.  
  Speed of sound dependence of the simulated Mach angles in the "`Hydro+Jet"' model, can not be reconciled with
theoretical estimate of Mach angles in a fluid with finite velocity.
The result agree with the finding in  \cite{Noronha:2008un}. In 
\cite{Noronha:2008un} it was shown that supersonic strings in
   AdS/CFT correspondence, do not produce observable conical angular correlation in strict $N_c \rightarrow  \infty$ supergravity limit. However, a special non-equilibrium neck zone, near the jet,   do produce Mach cone like peaks,   which does not follow Mach's law.

\section{summary and conclusions} \label{sec4}

Possibility of Mach shock wave formation  due to a quenching jet is studied as a function of the speed of sound of the medium. Modeling the quenching jet as a source, we have solved energy-momentum conservation equation of QGP fluid with a simple equation of state, $p=c_s^2 \varepsilon$, for $c_s^2$=0.1, 0.2, 0.3 and 0.4. Phase transition is neglected. It is assumed that at the freeze-out temperature $T_F$=100 MeV, QGP fluid instantly hadronizes.  In a single jet event,  simulated azimuthal distribution of pions due to a quenching jet does not resemble a conical flow. In a conical flow, a double peak structure, separated by twice the Mach angle ($\theta_M=cos^{-1}c_s$) is expected. Unlike in a conical flow, in a single jet event, simulated distribution is  single peaked, the peak position varying little with speed of sound. When averaged over all the jet trajectories, a double peaked structure, reminiscent of 'conical flow' emerges.   However, the peak positions are   at variance with that expected in a shock wave formation.   We conclude that the fine structure observed in experimental azimuthal distribution of secondaries associated with high $p_T$ trigger in Au+Au collisions  are possibly unrelated from Mach shock wave formation.


\begin{thebibliography}{99} 
 
\bibitem{QGP3jetqu} 
M. Gyulassy, I. Vitev, X.-N. Wang, and B.-W. Zhang,  
in {\it Quark-Gluon Plasma 3}, edited by R.~C. Hwa and  
X.-N. Wang (World Scientific, Singapore, 2004), p.~123. 
\bibitem{Adcox:2001jp}
  K.~Adcox {\it et al.}  [PHENIX Collaboration],
  Phys.\ Rev.\ Lett.\  {\bf 88}, 022301 (2002)
  [arXiv:nucl-ex/0109003].
\bibitem{Adler:2002xw}
  C.~Adler {\it et al.}  [STAR Collaboration],
  Phys.\ Rev.\ Lett.\  {\bf 89}, 202301 (2002)
  [arXiv:nucl-ex/0206011].
\bibitem{Adams:2003kv}
  J.~Adams {\it et al.}  [STAR Collaboration],
  Phys.\ Rev.\ Lett.\  {\bf 91}, 172302 (2003)
  [arXiv:nucl-ex/0305015].
\bibitem{Stoecker:2004qu}
  H.~Stoecker,
  Nucl.\ Phys.\ A {\bf 750}, 121 (2005)
  [arXiv:nucl-th/0406018].
\bibitem{shuryak} 
J.~Casalderrey-Solana, E.~V.~Shuryak and D.~Teaney, 
J. Phys. Conf. Ser. {\bf 27}, 22 (2005) [hep-ph/0411315]. 

\bibitem{Casalderrey-Solana:2005rf}
  J.~Casalderrey-Solana and E.~V.~Shuryak,
  arXiv:hep-ph/0511263.


\bibitem{Wang:2004kf}
  F.~Wang  [STAR Collaboration],
  J.\ Phys.\ G {\bf 30}, S1299 (2004)
  [arXiv:nucl-ex/0404010].
\bibitem{Jacak:2005af}
  B.~Jacak  [PHENIX Collaboration],
  arXiv:nucl-ex/0508036.
\bibitem{Dremin:2005an}
  I.~M.~Dremin,
  Nucl.\ Phys.\  A {\bf 767}, 233 (2006)
  [arXiv:hep-ph/0507167].
\bibitem{Koch:2005sx}
  V.~Koch, A.~Majumder and X.~N.~Wang,
  Phys.\ Rev.\ Lett.\  {\bf 96}, 172302 (2006)
  [arXiv:nucl-th/0507063].
\bibitem{Ma:2006mz}
  G.~L.~Ma {\it et al.},
  arXiv:nucl-th/0610088.
\bibitem{Chesler:2007an}
  P.~M.~Chesler and L.~G.~Yaffe,
  arXiv:0706.0368 [hep-th].


\bibitem{Renk:2005si}
  T.~Renk and J.~Ruppert,
  Phys.\ Rev.\  C {\bf 73}, 011901 (2006)
  [arXiv:hep-ph/0509036].

\bibitem{Ruppert:2007mm}
  J.~Ruppert and T.~Renk,
  Acta Phys.\ Polon.\ Supp.\  {\bf 1}, 633 (2008)
  [arXiv:0710.4124 [hep-ph]].

\bibitem{Torrieri:2009mv}
  G.~Torrieri, B.~Betz, J.~Noronha and M.~Gyulassy,
  Acta Phys.\ Polon.\  B {\bf 39}, 3281 (2008)
  [arXiv:0901.0230 [nucl-th]].
\bibitem{Betz:2008js}
  B.~Betz, M.~Gyulassy, D.~H.~Rischke, H.~Stocker and G.~Torrieri,
  J.\ Phys.\ G {\bf 35}, 104106 (2008)
  [arXiv:0804.4408 [hep-ph]].
 
\bibitem{Noronha:2008un}
  J.~Noronha, M.~Gyulassy and G.~Torrieri,
  Phys.\ Rev.\ Lett.\  {\bf 102}, 102301 (2009)
  [arXiv:0807.1038 [hep-ph]].
  
\bibitem{Betz:2008wy}
  B.~Betz, M.~Gyulassy, J.~Noronha and G.~Torrieri,
  arXiv:0807.4526 [hep-ph].

\bibitem{Neufeld:2008fi}
  R.~B.~Neufeld, B.~Muller and J.~Ruppert,
  Phys.\ Rev.\  C {\bf 78}, 041901 (2008)
  [arXiv:0802.2254 [hep-ph]].




\bibitem{Chaudhuri:2005vc}
  A.~K.~Chaudhuri and U.~Heinz,
Phys. Rev. Lett. {\bf 97}, 062301 (2006);
  arXiv:nucl-th/0503028.

\bibitem{Chaudhuri:2006qk}
  A.~K.~Chaudhuri,
  Phys.\ Rev.\  C {\bf 75}, 057902 (2007)
  [arXiv:nucl-th/0610121].
\bibitem{Chaudhuri:2008zz}
  A.~K.~Chaudhuri,
  Int.\ J.\ Mod.\ Phys.\  E {\bf 16}, 3131 (2008).
\bibitem{Chaudhuri:2007vc}
  A.~K.~Chaudhuri,
  Phys.\ Rev.\  C {\bf 77}, 027901 (2008)
  [arXiv:0706.3958 [nucl-th]].
 
\bibitem{Chaudhuri:2007gq}
  A.~K.~Chaudhuri,
  Phys.\ Lett.\  B {\bf 659}, 531 (2008)
  [arXiv:0705.1059 [nucl-th]].

\bibitem{Satarov:2005mv}
  L.~M.~Satarov, H.~Stoecker and I.~N.~Mishustin,
  Phys.\ Lett.\ B {\bf 627}, 64 (2005)
  [arXiv:hep-ph/0505245].
  

\bibitem{Eloss}
X.~N.~Wang,
Phys.\ Rev.\ C {\bf 70}, 031901 (2004), 
and private communication.

\bibitem{Baier:1996kr}
  R.~Baier, Y.~L.~Dokshitzer, A.~H.~Mueller, S.~Peigne and D.~Schiff,
  Nucl.\ Phys.\  B {\bf 483}, 291 (1997)
  [arXiv:hep-ph/9607355].
  
  
\bibitem{Gyulassy:2003mc}
  M.~Gyulassy, I.~Vitev, X.~N.~Wang and B.~W.~Zhang,
  arXiv:nucl-th/0302077.
  
\bibitem{Accardi:2004gp}
  A.~Accardi {\it et al.},
  arXiv:hep-ph/0310274.

\bibitem{QGP3v2} 
P.~F. Kolb and U. Heinz, in Ref.~\cite{QGP3jetqu}, p.~634. 
\bibitem{AZHYDRO}
P. F. Kolb, J. Sollfrank, and U. Heinz, Phys. Rev. C {\bf 62}, 054909 (2000);
P. F. Kolb and R. Rapp, Phys. Rev. C {\bf 67}, 044903 (2003). The code 
can be downloaded from URL http://nt3.phys.columbia.edu/people/molnard/OSCAR/


\bibitem{Back:2002wb}
  B.~B.~Back {\it et al.},
  Phys.\ Rev.\ Lett.\  {\bf 91}, 052303 (2003)
  [arXiv:nucl-ex/0210015].

\bibitem{Yarom:2007ni}
  A.~Yarom,
  Phys.\ Rev.\  D {\bf 75}, 105023 (2007)
  [arXiv:hep-th/0703095].
  
\bibitem{Chesler:2007sv}
  P.~M.~Chesler and L.~G.~Yaffe,
  Phys.\ Rev.\  D {\bf 78}, 045013 (2008)
  [arXiv:0712.0050 [hep-th]].
\bibitem{Neufeld:2008hs}
  R.~B.~Neufeld,
  Phys.\ Rev.\  D {\bf 78}, 085015 (2008)
  [arXiv:0805.0385 [hep-ph]].

\end{thebibliography}
\end{document}